\documentclass[12pt]{iopart} 

\usepackage{times}
\usepackage{enumerate}
\usepackage{color}
\usepackage{graphicx}
\usepackage{float}
\usepackage{mathrsfs}\let\rho=\varrho

\let\epsilon=\varepsilon

\let\epsilon=\varepsilon

\let\epsilon=\varepsilon

\def\sref#1{Sect.~\ref{#1}}
\def\fref#1{Fig.~\ref{#1}}
\def\ie{{\it i.e.}}

\def\b{{\rm burst}}
\def\p{{\rm isolated}}
\def\BB{{\cal B}}
\def\PP{{\cal P}}
\def\QQ{{\cal Q}}
\def\RR{{\cal R}}

\def\sref#1{Sect.~\ref{#1}}
\def\fref#1{Fig.~\ref{#1}}
\def\ie{{\it i.e.}}
\def\-{\kern -0.3em-\kern-0.3em}
\def\paragraph#1{\hfill\break{\it #1} }

\def\ie{{\it i.e.}}

\begin{document}

\title{Leadership in 2D living neural networks}

\author{J.-P. Eckmann$^{1,2}$,   Shimshon Jacobi$^3$,    Shimon Marom$^4$, Elisha Moses$^{3}$,
Cyrille Zbinden$^1$}

\address{$^1$ D\'{e}partement de Physique Th\'{e}orique, Universit\'{e} de
Gen\`{e}ve, CH-1211 Gen\`{e}ve 4, Switzerland.}
\address{$^2$ Section de Math\'{e}matiques, Universit\'{e} de
   Gen\`{e}ve, CH-1211 Gen\`{e}ve 4, Switzerland.}
\address{$^3$ Department of Physics of Complex Systems, The Weizmann Institute
of Science, Rehovot 76100, Israel.}
\address{$^4$ Department of Physiology and Biophysics, Faculty of Medicine, Technion, Haifa,
31096 Israel.}

\date{\today}

\begin{abstract}

Eytan and Marom \cite{MaromTopol} recently showed that the spontaneous burst
activity of rat neuron cultures includes `first to fire' cells that
consistently fire earlier than others. Here we analyze the behavior of these
neurons in long term recordings of spontaneous activity of rat hippocampal and
rat cortical neuron cultures from three different laboratories. We identify
precursor events that may either subside (`small events') or can lead to a
full-blown burst (`pre-bursts'). We find that the activation in the pre-burst
typically has a first neuron (`leader'), followed by a localized response in
its neighborhood. Locality is diminished in the bursts themselves. The long
term dynamics of the leaders is relatively robust, evolving with a half-life of
23-34 hours. Stimulation of the culture can temporarily alter the leader
distribution, but it returns to the previous distribution within about 1 hour.
We show that the leaders carry information about the identity of the burst, as
measured by the signature of the number of spikes per neuron in a burst. The
number of spikes from leaders in the first few spikes of a precursor event is
furthermore shown to be predictive with regard to the transition into a burst
(pre-burst versus small event). We conclude that the leaders play a r\^ole in
the development of the bursts, and conjecture that they are part of an
underlying sub-network that is excited first and then act as nucleation centers
for the burst.

\end{abstract}


\vspace*{5 mm} {\small \noindent {\it Keywords}: Neural activity, Living neural
networks, Burst initiation, Cortical and hippocampal neuron cultures.}

\vspace*{5 mm} {\small \noindent{{Corresponding author:} Elisha Moses. e-mail:
\color{blue} elisha.moses@weizmann.ac.il \color{black} }}

\submitto{\NJP} \maketitle

\section{Introduction}
Neurons that are extracted from the brain before they form connections can be
grown in a dish, where they proceed to build a random network that is different
from the one that they make within the brain \cite{MaromReview,PhysRep}.  As
this network matures it spontaneously develops an electrical activity that is
characterized by collective {\em bursts} \cite{TscherterInitiation,
MaedaGeneration1995, droge1986mac, PotterDevelopment}. The conceptual problems
underlying spontaneous excitation of the network as a whole involve both the
initiation and the propagation of the activity. Models of initiation usually
rely on the existence of a constant level of random firing in the network,
contributing to background noise that rises at times to a level that suffices
to excite the network.

When the neurons are restricted to grow along lines they create 1D cultures and
their firing pattern can be well understood in terms of a generation process
that occurs in one of several Burst Initiation Zones (BIZ), and a subsequent
propagation from that area into the rest of the line
\cite{feinerman2005,feinerman2007}. In 2D the situation is much less clear.
While numerous studies identify BIZ's in 2D as well, the course of the outward
propagation is not concentric and is difficult to follow \cite{VanPeltDynamics,
TscherterInitiation, Darbon2002, YvonSpont}. Since signals are confined to
propagate along the axons that are linear and not space filling, advance of the
signal in the plane must be the collective effect of many neurons.

Eytan and Marom \cite{MaromTopol} recently found that in 2D a select group of
neurons, termed `first-to-fire', or `precursor' neurons, begin to fire ahead of
the rest of the network, and were seen to do so consistently for the several
hours that they were monitored. This raises several interesting questions
regarding their r\^ole in activating the network. Are these neurons simply
better connected to the actual initiation zone, or do they actively participate
in creating and propagating the burst. Since the experiments are performed with
multi electrode arrays that typically monitor only about $0.1\%$ of the neurons
in the culture, a further question is, how many such 'first to fire' groups
exist in the whole culture and how do they interact with each other. The
possibility of a `sub-network' of such neurons that ignites the rest of the
culture is an intriguing one.

In this study we explore the collective mechanism for the initiation of the
burst. Our approach will be to look first into the individual structure of each
burst and to study the properties of the `first-to-fire' neurons. We then ask
whether and how their activity actually predicts properties of the burst that
follows.

\section{Methods} \label{s:methods}

\subsection{Neural Culture Types}

We analyzed data from two-dimensional cultures prepared in three different
protocols taken in three different labs. They are classified according to their
respective feeding protocols, which we found to be the most significantly
differentiating parameter of the growth and preparation conditions:

\begin{itemize}
     \item \textit{`CF'} for a Continously-Fed culture, prepared from cortices
of post-natal rats that are perfused \cite{MaromTopol}, from the Marom lab.
     \item \textit{`DF'} for Daily Fed cultures, prepared from hippocampi of
pre-natal E19 rats, from the Moses lab.
     \item \textit{`WF'} for cultures fed every 5-7 days, prepared from
cortices of E18 rats \cite{PotterDevelopment}, from the Potter lab.
\end{itemize}

The three culture types are compared in Table \ref{tab:CultureTypesTable}
below.

\begin{table}[h]
\begin{center}
\small
     \begin{tabular}{ | l | p{2.0cm} | p{2.5cm} | p{1.35cm} | p{1.25cm} | p{2cm}
| p{1.3cm} |}
     \hline
     Name & Feeding Schedule & Source & Culture diameter [mm] & Feeding Medium &
Array size, $X \times Y$ [mm] and D [$\mu$m] & no.~of cultures \\ \hline
     \textit{CF} \cite{MaromTopol} & continuous perfusion from DIV 14 & Rat
cortical, postnatal day 1
         & 15 & 5\% horse serum & $1.4 \times
  1.4$ and 200 & 1 \\ \hline
     \textit{DF} & 33\% medium replacement, daily & Rat hippocampal, prenatal day 19
         & 15 & 10\% horse serum &  $2.5 \times
  4.5$ and 500 & 9 \\ \hline
     \textit{WF} \cite{PotterDevelopment} & 50\% medium replacement, every 5 days &
     Rat cortical, prenatal day 18  & 4 & 10\% horse serum & $1.4
  \times 1.4$ and 200 & 4 \\
     \hline
     \end{tabular}
\end{center}
\caption{\small Culture types used in this work. $X \times Y$ are the array
dimensions, D is the inter-electrode distance, and DIV stands for days
in-vitro.} \label{tab:CultureTypesTable} \small
%
%
%

\end{table}

\subsection{Preparation and Measurement of \textit{DF} Cultures} Here we only
describe the preparation of \textit{DF} cultures. The preparation of
\textit{CF} is described in \cite{MaromTopol}, and the preparation of
\textit{WF} cultures in \cite{PotterDevelopment}.

\paragraph{MEA Preparation.}Multi-electrode arrays (MEA) of $6\times10$
electrodes with $500 \mu m$ spacing were used (MultiChannelSystems, Reutlingen,
Germany). The MEA were prepared using overnight application of PEI solution
(Poly Ethylene Imine, Sigma, 0.05\% in borate buffer). 2 hours prior to cell
plating, the MEA was washed 4 times with double distilled water, and left with
plating medium.

\paragraph{Culture Preparation and Maintenance.} Rat hippocampal
neurons were taken from embryos of 19 day pregnant Wistar rats. The dissection
and plating were done according to
\cite{PapaSegalMorphologicalPlasticity,jacobi2007}, with daily feeding of the
culture. Briefly, the hippocampus was mechanically dissociated, and cells were
plated in a concentration of $12000$ cells$/$mm$^{2}$ in 3 ml plating medium
(5\% fetal calf serum (FCS, Sigma) and 5\% heat inactivated horse serum (HIHS,
Sigma) in Eagle's MEM (Sigma) enriched with 0.6\% glucose, 2 mM glutamine, and
$15$mg$/$ml Gentamicin (MEM+3g), enriched with B-27 supplement). In day
in-vitro (DIV) 3--6 the culture was fed daily with 1ml of changing medium
which is comprised of MEM+3g, 5\% FCS, 5\% HIHS with 20mg$/$ml FUDR and 50mg/ml
Uridine (both from Sigma), and from then on it was fed with 1ml final medium
which includes MEM+3g with 10\% HIHS. The resulting neuron density is about
$2000$ cells/mm$^{2}$.

\paragraph{Culture Measurement.}
Measurements were performed at 37$^{\circ}$C, in a dry 5\% CO$_{2}$ incubator.
To minimize evaporation of the growing medium, the culture chambers were
covered with a thin teflon foil \cite{potter2001nan}. For each culture, we
measured regularly and in a continuous way the spiking activity every 24 hours,
in DIV 5--35. The measurements for each culture were performed over times
ranging between 1 to 24 hours, from which extracts lasting 20 to 30 minutes
were analyzed. Measurements which occurred less than 15 minutes after moving
the culture or less than 6 hours since the last feeding were discarded. Special
care was taken to assure minimization of water loss from the growing medium
(\cite{jacobi2007,potter2001}).

\paragraph{Spike Detection.}
The detection was done after \cite{jacobi2007}. The output from the MEA1060
amplifier (MultiChannelSystems, Germany) was sampled at 20kHz (PCI-6071E,
National Instruments, Austin, TX). The condition for spike detection was that
the absolute value of the sampled signal exceeds the threshold for at least
0.2ms. The threshold was set at the maximum of $15$ $\mu V$ and 3 times the
signal standard deviation (which is $\sim3.5$ $\mu V$). Further spikes with
opposite polarity and less than 3ms apart were disregarded as they were usually
caused by overshoot. Spike shape discrimination was not performed, so that the
analysis relies only on the spike timing and the electrode position. The number
of neurons recorded from each electrode was estimated separately by spikes
sorting as 1--4, with an average of 2.5.

\paragraph{Culture Stimulation.}
We used stimulation to modify the leader distribution, in two methods. The
first method used one of the recording electrodes, to which a 0.6--1 Volts
amplitude, $1$ms bipolar pulse pair with $10$ms inter-pulse interval was
injected. The second method was bath stimulation, where two platinum electrodes
were immersed in the growing medium $4$mm aside from the electrode array
center. The bath stimulation was not carried out directly over the electrode
array to minimize the measurement dead time caused by the stimulation current.
The electrodes were exposed for $10$mm, spaced $2.5$mm apart and were less than
1mm above the neuron level. In this case, bipolar pulses with an amplitude of
2--3.5 Volts and width of $20$ms were used.

For both methods we used a battery powered current source which is isolated
from the surrounding environment to minimize electrical noise in the
multi-electrode measurement. The culture was monitored in order to verify its
response to the stimulation. The voltage of the stimulation was adjusted to
cause a culture-wide response in at least 50\% of the stimulations when the
stimulation frequency was set to the spontaneous bursting frequency.

\subsection{Data Analysis} \label{s:analysis}

\subsubsection{Time course}
Cultures were observed in days in-vitro 5--30 or parts within.
Measurements were taken typically about once a day, on the order of
an hour. Each daily data set is termed \emph{epoch}. There were on
average 8, 25 and 20 epochs in \textit{CF}, \textit{DF},
\textit{WF}, respectively. We looked for changes of behavior across
epochs, typically finding changes towards the middle of the
measurement period. This naturally leads to a division of the epochs
into two \emph{superepochs}. Electrodes were included in the
analysis of a given epoch (`active electrode') if they fired at
least once in this epoch.

\subsubsection{Burst definition}
The definition of bursts is based on the list of spikes which are
obtained in the measurements. For each spike $i$, the list gives us
$t_i$, the time of the $i$th spike, and $c_i$, the electrode
(`\emph{channel}') of that spike.

\begin{figure}[!h]
\begin{center}
\includegraphics[width=0.6\textwidth,angle=0]{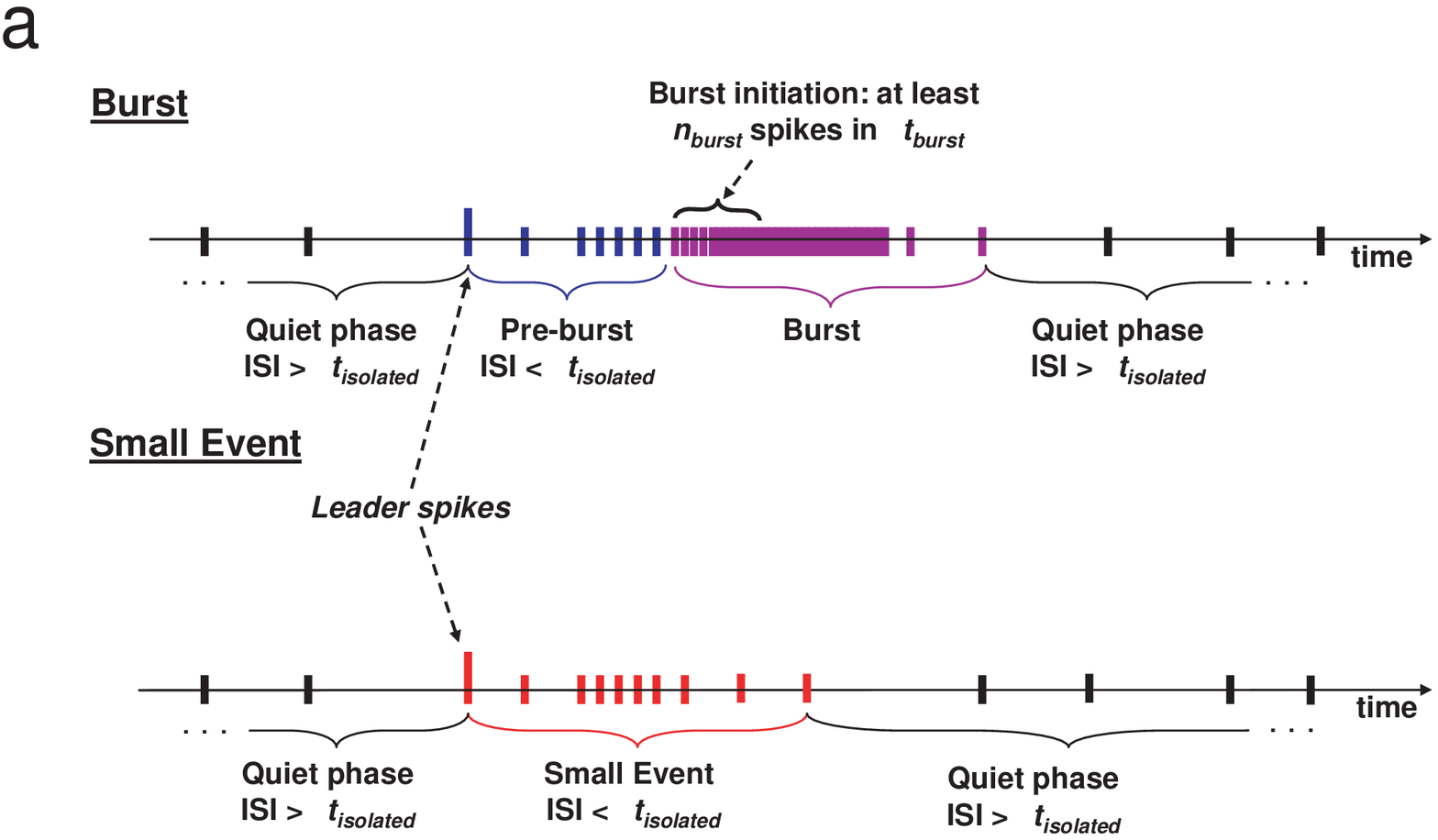} 
\includegraphics[width=0.6\textwidth,angle=0]{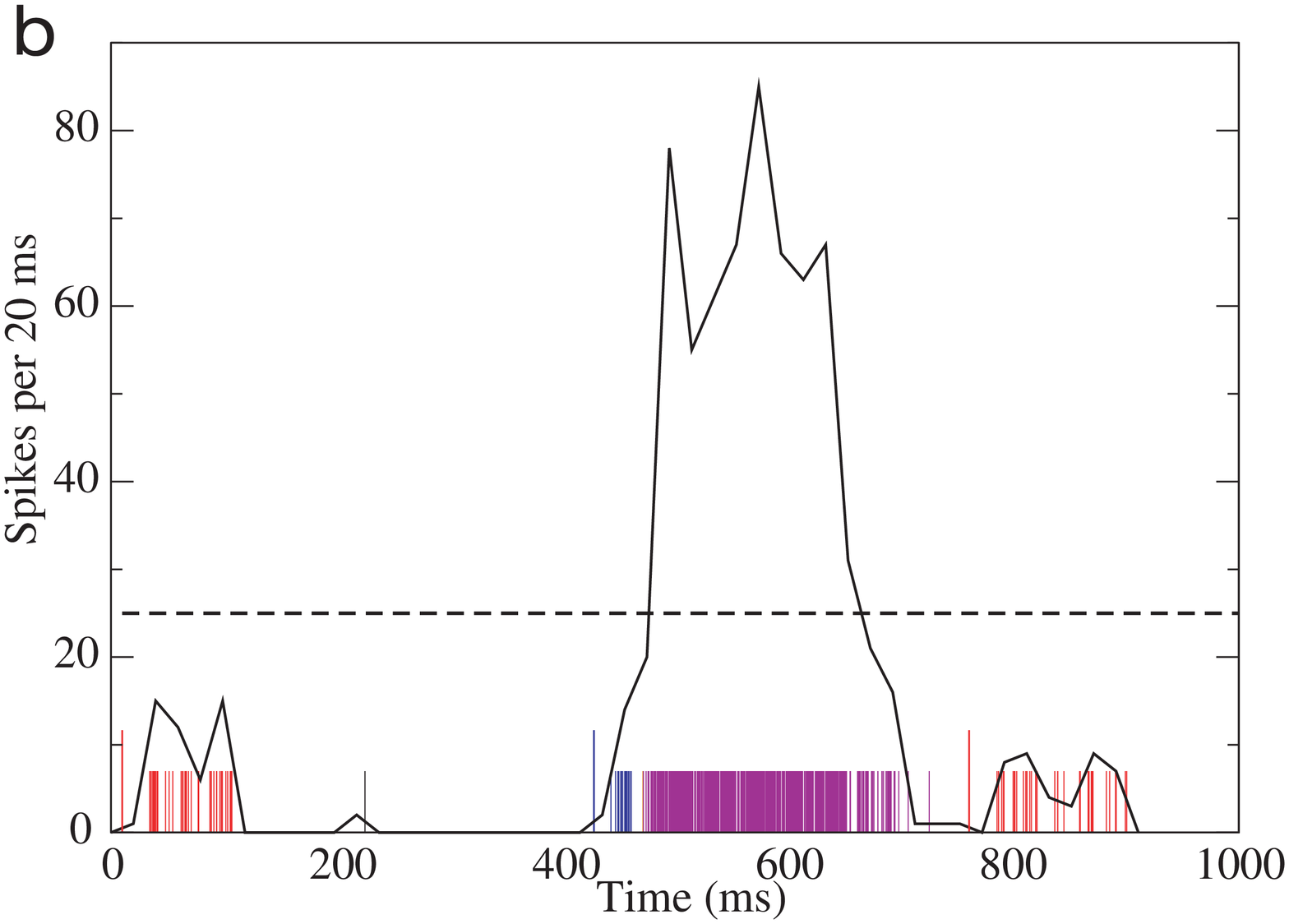}
\caption{Burst identification. a. A Schematic drawing of the classification of
spikes. The verticals line indicate spikes, with the following color-code:
Violet (burst), blue (pre-burst), red (small event), black (isolated events).
Spikes with inter spike interval below $\delta t _{isolated}$ are grouped into
a pre-burst that is followed by a burst if the spike density exceeds
$n_{burst}$ spikes in $\delta t _{burst}$ (top timeline), or into a small event
otherwise (bottom timeline). b. Example of spontaneous bursting activity in a
\textit{CF} culture. Single spikes are depicted as vertical lines with the
color-code as in a, while the number of spikes per 20ms bins are drawn as a
line. This data set includes one {\em isolated spike}, two {\em small events}
and one {\em burst}. The leader spikes are shown with a somewhat longer
vertical line. The dashed line is the threshold value $n_\b=25$ in $\delta t
_\b=20$ms for \textit{CF}.} \label{f:sample}
\end{center}
\end{figure}

We will associate each spike with one of 4 classes: \emph{in-burst},
\emph{pre-burst}, \emph{small-event}, \emph{isolated}. Basically, a
spike is in a burst if it is in a group of many spikes which follow
each other closely. It is in a pre-burst or small-event if it is not
in a burst and is in a sequence of spikes close enough in time so
that communication between them is still possible. The distinction
between pre-burst or small event depends on whether the spike is
eventually followed by a burst or not. Finally, all other spikes are
isolated. The reader may want to refer to \fref{f:sample} for the
definitions which follow.

More precisely, we use three parameters, $n_\b$, $\delta t_\b$, and
$\delta t_\p$. We first define consecutive sets, separated by
interspike gaps of length at least $\delta t_\p$. Thus the set of
spikes is divided into sets  $\RR_\ell$ of consecutive spikes, with
the property that if $i$ and $i+1$ are in $\RR_\ell$ then
$t_{i+1}-t_i \le \delta t_\p$, while for $i\in\RR_\ell$ and $i+1\in
\RR_{\ell+1}$ we have $t_{i+1}-t_{i}>\delta t_\p$. The rationale is
that the interspike gap is large enough so that no `memory' remains
between the spike at $t_i$ and that at $t_{i+1}$ when
$t_{i+1}-t_{i}>\delta t_\p$.

Each of the $\RR_\ell$ contains at least one spike, but may contain many. We
subdivide each $\RR_\ell$ into 2 disjoint sets $\RR_\ell=\PP_\ell^{(0)}\cup
\BB_\ell^{(0)}$ using the following condition: We first define $i_*=i_*(\ell)$
to be the first index in $\RR_\ell$ with the property:
$t_{i_*+n_\b-1}-t_{i_*}<\delta t_\b$, that is, there are at least $n_\b$ spikes
in $\RR_\ell$ within a lapse of time $\delta t_\b$. If the condition is never
met, then the set $\RR_\ell$ is not subdivided and is called a \emph{small
event} if it has more than 1 spike and is an \emph{isolated spike} otherwise.
For those cases where such an $i_*$ can be found, we check if it is also the
first spike in the burst. If so, then that burst is an \emph{immediate burst}
that has no \emph{pre-burst}, and $\RR_\ell=\BB_\ell^{(0)}$. For all the other
bursts (these are the majority) we divide $\RR_\ell=\PP_\ell^{(0)}\cup
\BB_\ell^{(0)}$ where $\PP_\ell^{(0)}$ are the indices $i\in\RR_\ell$ with $i <
i_*$, and $\BB_\ell^{(0)}$ are the others. The letters $\PP$ and $\BB$ refer to
pre-burst and burst.

We now renumber the set of indices as follows
$$
\{1,\dots,N\} = \QQ_1 \cup \PP_1 \cup \BB_1\cup
\QQ_2\cup \PP_2\cup \BB_2 \cup \dots~,
$$
where each of the $\PP$ and $\BB$ equals one of the earlier $\PP^{(0)}$ and
$\BB^{(0)}$ but the $\QQ$ are empty sets or unions of consecutive $\RR$'s which
were not identified as bursts in the earlier stage.

We call $\BB_\ell$ the \emph{burst} number $\ell$, $\PP_\ell$ the corresponding
\emph{pre-burst} and $\QQ_\ell$ the corresponding \emph{quiet phase}. The
\emph{leader} of burst $\ell$ is the channel $c_{i_\ell}$, where $i_\ell$ is
the number of the first spike in $\PP_\ell$.

Note that it may happen that $\QQ_\ell$ does not contain any spike, but we
always can define a quiet duration. This duration is the time between the last
spike of $\BB_{\ell-1}$ and the first spike of $\PP_\ell$. We define $T$ as the
combined length of the quiet phases. The length of the quiet phase $\QQ_\ell$
is $t_{i_{\ell}}-t_{\widehat{i}_{\ell-1}}\ (>\delta t_{\p})$ where $i_\ell$ is
the first spike in $\PP_\ell$ (\ie, the leader of burst $\ell$) and
$\widehat{i}_{\ell-1}$ is the last spike of $\BB_{\ell-1}$. Therefore,
$$
T=\sum_{\ell}t_{i_{\ell}}-t_{\widehat{i}_{\ell-1}}\,,
$$
is the total duration of all quiet phases.

The parameters which were used for the 3 cultures are shown in Table
\ref{tab:BurstParamTable}.

\begin{table}[h] 
\begin{center}
\begin{tabular}{|c|c|c|c|}
   \hline
   Culture Type & $n_\b$ & $\delta t_\b$ [ms] & $\delta t_{\p}$ [ms] \\
   \hline
   \textit{CF} & 25 & 20 & 20 \\
   \textit{DF} & 10 & 40 & 50 \\
   \textit{WF} & 20 & 20 & 10 \\
   \hline
\end{tabular}
\end{center}
\caption{Burst detection parameters.} \label{tab:BurstParamTable}
\end{table}

With these definitions we can do now some statistics: We are
interested in the activity of the system \emph{outside} the bursts.
The question is whether certain channels are the leaders of the
neural activity more often than warranted by simple expectation.

So we start by estimating the expected rate (or spiking density) for
each channel. Two choices can be used here: Averages over the
$\QQ_\ell$ or averages over $\QQ_\ell$ and $\PP_\ell$. Let us continue
with the first option, this is the one used in the experiments.

For each channel $n$, we define $C_n$ as the number of times $n$ has
spiked in the set
$$
\QQ = \QQ_1 \cup
\QQ_2\cup \dots ~.
$$
The rationale for this is that some channels spike only in the bursts. So the
channels that are able to be leaders are those which can spike in the quiet
phase. However, we have checked that for most channels the activity in the
quiet phase is comparable to that in the bursts (data not shown).

\paragraph{Leader Significance.}
For every pre-burst, small event or immediate burst, the electrode
which fires first is called a \emph{leader}. Since we are interested
in their special r\^ole, we first need to make sure that leaders are
not just electrodes with high activity, which therefore are
statistically more often the first ones to fire. The following
discussion will show that they are over-proportionally more often
leaders.

Henceforth we only discuss bursts with a pre-burst and their
\textit{leaders}. We require that a leader's probability to lead
bursts should be significantly higher than its probability to fire
in low-rate (quiet) intervals. Let $M$ be the total number of bursts
that have a pre-burst, which is also the total number of burst
leaders. We consider a spike in the quiet interval, and term by
$q_n$ the probability that electrode $n$ has fired that spike. The
a-priori probability $P_n(k)$ that the channel $n$ is $k$ times the
leader is then
\begin{equation*}
   {M \choose k} q_n^{k}(1-q_n)^{M-k}~.
\end{equation*}
In the limit of large $M$ and reasonable $q_n$ this distribution is
approximated by a Gaussian of mean $Mq_n$ and variance $Mq_n(1-q_n)$. We denote
by $B_n$ the \emph{actual} number of bursts with a pre-burst that electrode $n$
leads (note that $\sum_n {B_n} = M$). Thus we have a scale on which to test
leadership. We define $\alpha _n$, a `leadership score', and decide that an
electrode is a \emph{significant leader} if it is at least 3 standard
deviations above the natural expectation value. This corresponds to a p-value
of $0.001$. The criterion for significant leadership is therefore
\begin{equation}
\alpha _n=\frac{B_{n}-Mq_{n}}{\sqrt{Mq_{n}(1-q_{n})}}  ~>~ 3~.
\end{equation}
A second condition is aimed to eliminate electrodes that fire very seldom and
lead a negligible number of bursts but are not screened by the first condition.
In other words, an electrode cannot be a leader if the fraction of bursts it
leads is less than a threshold of $3\%$ of the bursts.

\paragraph{Mutual Information.}
We estimate the mutual information of two electrodes in a series of
time intervals based on empirical probabilities, according to
\cite{Email}. Using the division of time to pre-burst and burst
intervals, we define a firing event in a given interval as $k=0,1$,
where 0 stands for no spike and 1 stands for at least one spike. We
did not use the quiet interval for this analysis. We denote by
$N_n(k)$ the number of times electrode $n$ had an event $k$ in a
series of $N_{\rm int}$ intervals. The probability assigned to event
$k$ is thus $p_n(k)=N_n(k)/N_{\rm int}$. The number of joint events
in which electrode $n$ had event $i$ and electrode $n'$ had event
$j$ is given by $N_{n,n'}(i,j)$. The joint probability of events
$i,j$ for electrodes $n$,$n'$ is then given by
$p_{n,n'}(i,j)=N_{n,n'}(i,j)/N_{\rm int}$. The mutual information
between two electrodes is then given by

\begin{equation*}
I_{n,n'}=\displaystyle\sum_{i,j=0,1}p_{n,n'}(i,j)\cdot
\log\displaystyle\frac{p_{n,n'}(i,j)}{p_n(i)p_{n'}(j)}
\end{equation*}



\section{Results}

\subsection{Neural Activity}

All three culture types showed spontaneous bursting in most of the active
electrodes. In the cultures from all sources \textit{CF} (a single culture),
\textit{DF} (n=9 cultures), \textit{WF} (n=5 cultures) the average rate of
spikes per electrode per unit time gradually increased along the measurement
period as did the bursting rate. The measurement period was (in DIV) 14--25
(\textit{CF}), 5--30 (\textit{DF}), and 5--30 (\textit{WF}). The electrode
firing rate (in Hz per active electrode) was 3--10 (\textit{CF}), 1--3
(\textit{DF}) and 4--10 (\textit{WF}). For all cultures more than $85\%$ of
the intact electrodes were firing together in bursts after DIV 14.

The main difference between the activity of the cultures lies in the
rate of firing. In particular, maximal bursting rates were (in Hz)
$1.6$ (\textit{CF}), 0.3--1.4, with an average of $0.65$
(\textit{DF}), and 0.07--0.84, with an average of $0.26$
(\textit{WF}). These maximal rates were obtained in ages of DIV
19--24 (\textit{CF}), 14--21 (\textit{DF}), and 21--28
(\textit{WF}). The difference between the bursting rates can mainly
be attributed to the feeding differences - we have seen that more
frequent feeding increases the bursting rate \cite{jacobi2007}. By
the third week-in-vitro of the culture, about $90\%$ of the spikes
were within bursts for all three culture types.

\textit{WF} cultures were special in that they contained fewer
bursts but were more active in the quiet intervals between the
bursts, with an average firing rate (excluding bursts) of about 0.1
Hz per electrode as compared to 0.002 Hz (\textit{DF}) and 0.02 Hz
(\textit{CF}).

\subsection{Existence of Leaders} Most of the bursts originated in a
pre-burst (94\% for \textit{CF}, 86\% for \textit{DF}, and 73\% for
\textit{WF} cultures).

The pre-burst lasted on average $230$ms in \textit{DF}, $180$ms for
\textit{CF} and $22$ms for \textit{WF}. The average number of
participating electrodes was $10$ in \textit{DF}, $18$ in
\textit{CF} and $6$ in \textit{WF}. The \textit{bursts} lasted on
average $500$ms in \textit{DF}, $314$ms for \textit{CF} and $228$ms
for \textit{WF}. The corresponding average number of firing
electrodes was $29$ in \textit{DF}, $48$ in \textit{CF} and $31$ in
\textit{WF}.

In all epochs, only $4\pm1.9$ (mean$\pm$SD), $2.9\pm1.2$, and $2.0\pm1.4$
electrodes are significant leaders for more than 24 hours in \textit{CF},
\textit{DF} and \textit{WF} cultures, respectively.

\subsection{Time evolution}
\begin{figure}[!h]
     \begin{center}
\includegraphics[width=0.72\textwidth,angle=0]{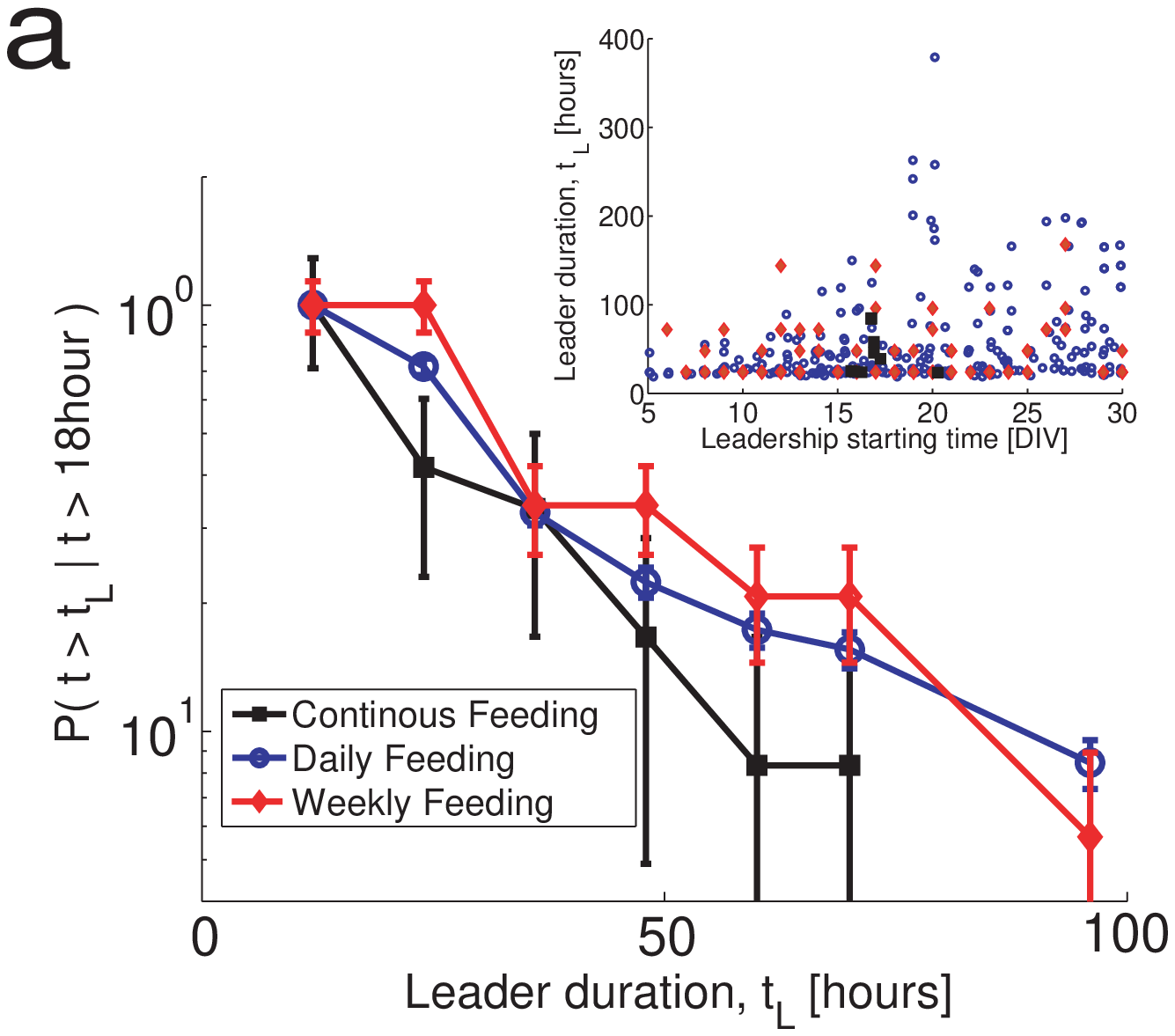} %
\includegraphics[width=0.6\textwidth,angle=0]{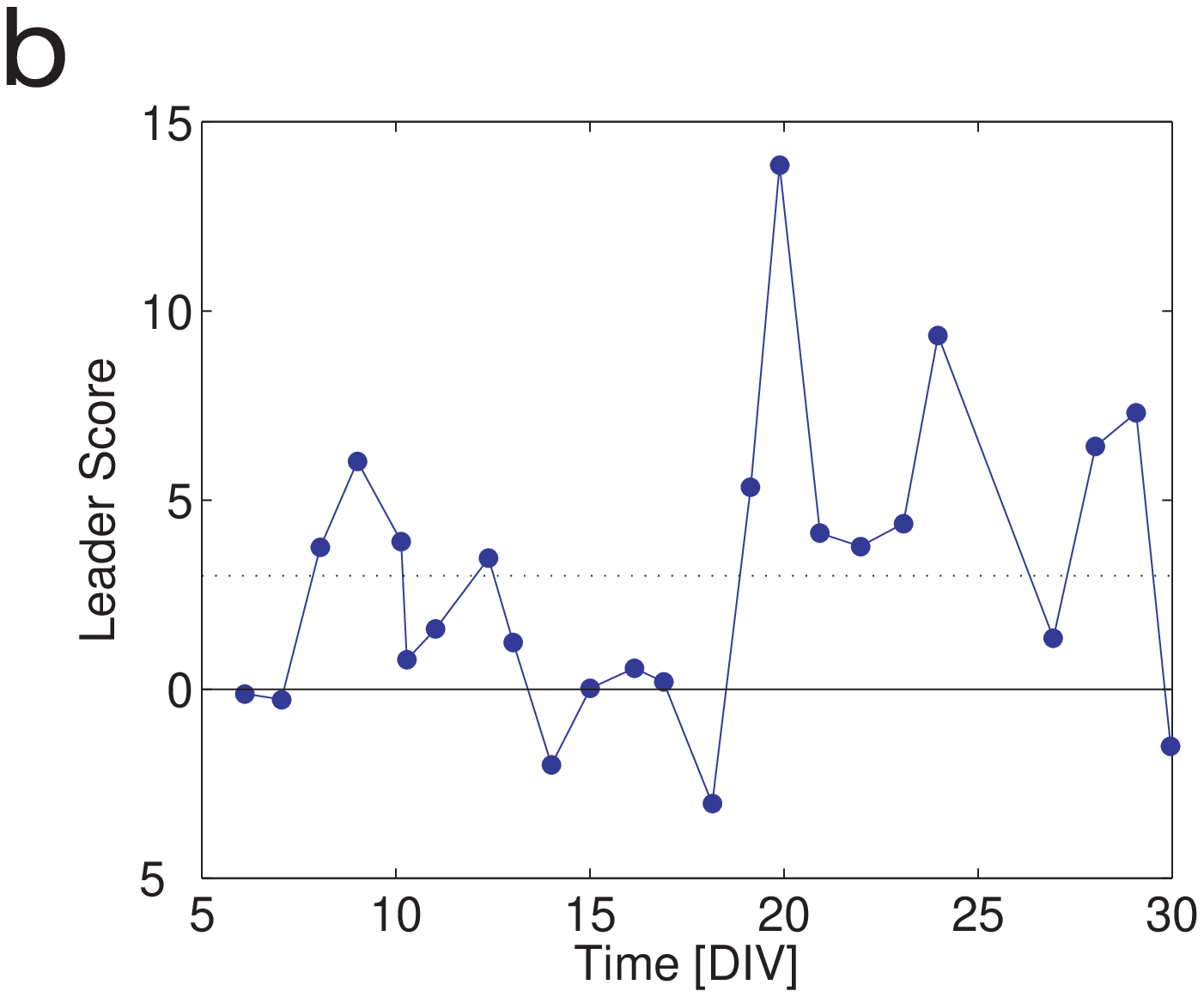} %
\caption{Leader electrodes remain with a high leadership score for days. a. The
distribution of the leadership duration, for the three culture types,
\textit{CF}, with continuous feeding (black squares), \textit{DF}, with daily
feeding (blue circles), \textit{WF}, with semiweekly feeding (red diamonds).
Inset: leadership time durations of all leaders remaining at least 18 hours, as
a function of the leadership starting time (marker code as in a.). The b.
Leadership score as a function of day \textit{in-vitro}, for an example leader
electrode of \textit{DF}. } \label{f:LeaderStability}
     \end{center}
   \end{figure}

We are interested in finding out whether leadership is maintained over
successive epochs. It turns out, that for all cultures, the most significant
leaders are indeed stable over multiple epochs (\fref{f:LeaderStability}). To
quantify the leadership inertia, we define $L_{n}$ as the set of epochs in
which electrode $n$ is a significant leader, \ie,
\begin{equation}
   L_{n}=\{k~|~ \alpha_{\rm c}<\alpha_{n,k}\}~,
\end{equation}
where $\alpha_{\rm c}=3$ is the threshold leadership score, and $\alpha_{n,k}$
is the leader score for electrode $n$ in epoch $k$. For every consecutive
sequence $k,k+1,\dots,\ell$ of integers in $L_{n}$, we define the lifetime of
leadership by $t_{\ell+1}-t_k$.

The \emph{distribution} of the lengths of all these time intervals follows an
exponential distribution with half-life (in hours) of $23.1\pm3.0$
(mean$\pm$SE) (\textit{CF}), $34.4\pm4.4$ (\textit{DF}), and $29.5\pm3.6$
(\textit{WF}).

The difference between the culture types is probably caused by the culture
maintenance protocol.

\subsection{Locality of activity}
\begin{figure}[!h]
     \begin{center}
\includegraphics[width=0.6\textwidth,angle=0]{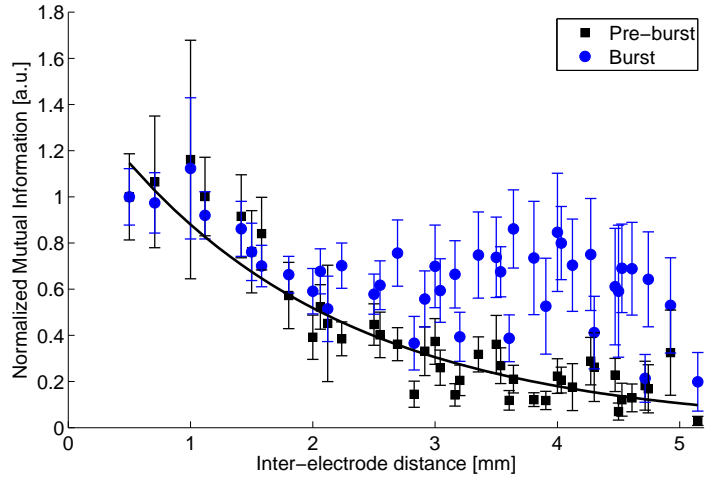} %
\caption{Average Mutual Information (MI) as a function of distance
for pre-burst and burst intervals. The MI is normalized by the its
value at 0.5mm. The activity landscape is more localized in the
pre-burst, and can be fit to an exponential (black line) with length
scale of $1.3$mm. The length scale in the burst is higher (blue line
is a guide to the eye). }
       \label{f:two-dim-locality-picture}
     \end{center}
   \end{figure}

We assess the locality of electrode co-activation within the same
burst or pre-burst, by measuring the Mutual Information as a
function of distance. We compare the pre-burst to the burst, and
show that activity in the pre-burst is localized while the activity
in the burst tends to spread all over the culture. We quantify the
tendency to fire together by estimating the mutual information
between electrode pairs (see \sref{s:methods}). A pair of electrodes
has higher mutual information if they participate in the same
pre-burst or burst interval. We plot the mutual information
$I_{n,n'}$ as a function of $d_{n,n'}$, the distance between the
electrodes $n$ and $n'$, averaging over all electrode pairs which
have the same distance. The mutual information decreases with the
inter-electrode distance for all interval types. In the pre-burst,
we get a reasonable exponential fit of the form $I(d)=A \cdot \exp (
-d/L )$ (for example, \fref{f:two-dim-locality-picture}).

The mutual information displayed in \fref{f:two-dim-locality-picture} is
normalized so that a direct comparison can be made between the burst and the
pre-burst. However, the actual values of the measured information are also of
interest. The mutual information in the burst is typically 5 times higher than
in the pre-burst. This shows that in the burst, the co-activation is much
higher than in the pre-burst. Looking at \fref{f:two-dim-locality-picture} one
can see that in the pre-burst the mutual information decays gradually, going to
zero at long distances. In contrast, during the burst it has a short initial
decay, on the order of 1--2mm, beyond which the information is maintained at a
constant value. We conclude that in the pre-burst the activity is primarily
local, decaying to zero within $5$mm. In contrast, in the burst the local
effect is limited (within 1--2mm), and once the burst develops and advances
outside this local region then all the culture behaves as one correlated
region.

The value of $L$ in the pre-burst interval can be measured reliably in epochs
of maximal activity, occurring in DIV 19--23 (\textit{CF}), 10--20
(\textit{DF}) and 15--24 (\textit{WF}). At other times, the decay is too small
to calculate $L$ reliably. In these days, $L$ (in mm) of the pre-burst interval
either wanders between 0.64--1.8 as function of culture age (\textit{CF}), or
takes constant values of $1.85\pm0.16$ (\textit{DF}), $1.09\pm0.02$
(\textit{WF}). In all epochs of maximal activity defined above, the
localization in the burst interval as well as in the quiet phase is weaker than
in the pre-burst interval.

As it is known that the recruitment of neural activity in bursts is based on
synaptic transmission \cite{MaromReview,MaedaGeneration1995}, we relate the
correlation length to the number of synaptic connections that the signal must
propagate through. This in turn is determined by the axonal length, which is
also in the range of about 1mm in similar cultures
\cite{Bartlett1984ems,Kriegstein1983Morphological}.

\subsection{Effect of stimulation}

We have seen that the leaders can remain stable over long times when the
culture is spontaneously active. We wished to assess whether the leaders can be
changed using an external driving force \cite{MaedaModification}. In order to
modify the leader distribution we stimulated the cultures using either one of
the MEA electrodes or bath electrodes (see \sref{s:methods}). These experiments
were only done on two \textit{DF} cultures, additional to the 9 cultures
discussed above. The stimulation was repeated 40--240 times, in stimulation
frequencies of 0.2--1 Hz during periods of 2--5 minutes for bath stimulation,
and 500--1000 times in frequencies of 0.1--0.3 Hz during periods of 20--30
minutes for single-electrode stimulation. The response of the culture to the
stimulation did not exceed the spontaneous bursting rate: stimulation in higher
frequencies resulted in a lower fraction of bursts that did occur as a response
to stimulation.

We assess the leader distribution measurement resolution by considering a
combined all-channel criterion for measuring the modification in the leader
distribution that is based on $B_n$, the number of bursts lead by electrode
$n$. We define a similarity measure between two leader distributions ${\bf B}'
= \{B'_n\}$ and ${\bf B} = \{B_n\}$ as their normalized scalar
product, 
\begin{equation}
S = \frac {{\bf B'} \cdot \bf B} { |\bf B'| \cdot |\bf B|}~.
\end{equation}

The similarity $S$ is bounded between $0$ and $1$, where $S=1$ means
identical leader distributions while a similarity of $0$ means that
all the leaders have been completely replaced by new leader
electrodes. The reference distribution $B'_n$ was measured during a
30--60-minute interval which ends 15 minutes before the
stimulation, and for $B_n$ we used a sliding 10-minute interval
either before (t$<$0) or after (t$>$0) the stimulation. Without
stimulation, $S$ slowly wanders due to distribution's dynamics,
starting close to $1$ and approaching values of 0.6--0.9 after $2$
hours. To allow a clear comparison, we normalize the value of $S$ by
$S_{\rm ref}$, its mean value during the 15 minutes prior to the
stimulation.

\begin{figure}[!h]
     \begin{center}
        \includegraphics[width=0.66\textwidth,angle=0]{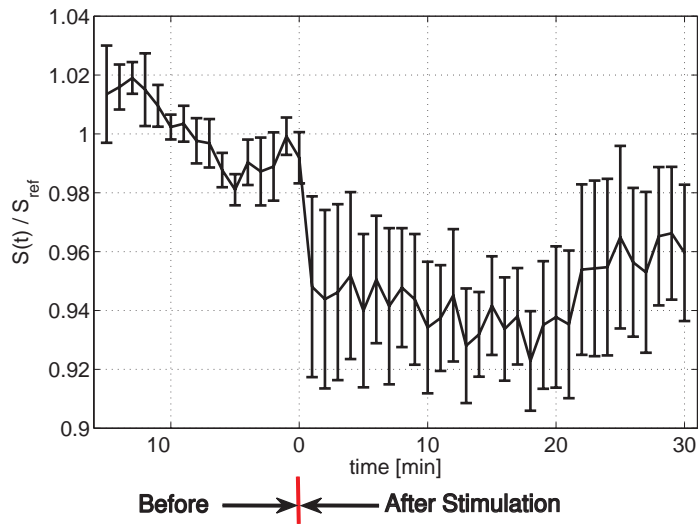} %
\caption{Stimulation drives changes in Leader distribution, that fade out
within few tens of minutes. S(t) is the similarity between the leader
distributions of 30--60min pre-stimulation, and a sliding 10min window either
pre-stimulation (t$<$0) or post-stimulation (t$>$0). The value of S is
normalized by its average during the 15 minutes prior to stimulation. Vertical
red line border between pre- and post-stimulation regions: note that the
stimulation period, of 2-10min, is not present in the graph. S(t) declines
immediately following the burst and climbs up after about 30min. }
       \label{f:StimulationSummary}
     \end{center}
   \end{figure}

Bath stimulation results are presented in \fref{f:StimulationSummary} for
cultures in age of 12--14 DIV. After the stimulation, the deviation rises
abruptly, meaning that the short-term stimulation effect is maximal. The
variation between experiments causes the large spread of results which is
visible in short time after the stimulation. After 20--30 minutes, the
similarity climbs back towards $1$. In younger cultures aging 6--8 DIV, the
average decline in similarity is three times higher, the variation between
experiments is also three times higher, and the return to $1$ somewhat faster
(data not shown). The climb of similarity towards $1$ indicates that the impact
of the stimulation is fading. We note that opposite to the bath stimulation
case, single electrode stimulation did not result in any modification in the
leader distribution (results not shown).

We conclude that the bath stimulation effects are limited to less
than an hour, and that the leader distributions of more mature
cultures show lower susceptibility to stimulation.

\subsection{Predictability of activity}

The leader electrodes can be, on one hand, sensitive probes of the
average whole-culture activity, or, on the other hand, be the areas
that are leading the recruitment of the rest of the culture. In
order to judge between these alternatives, we have been able to
identify properties of the burst that correlate with the identity of
the leader. In the following paragraphs, we show that the identity
of the leader affects the activity following it in both the
pre-burst and the burst intervals.

To test the relation between the leaders and the bursts, we analyzed the data
in two ways. First, we determine a small number $n$ of leaders, and for every
burst initiated by that leader, we determine the number of spikes in each
channel. Thus, we have a sequence of vectors ${\bf c}_{\ell,i}$, where the
$j$'th element of the vector ${\bf c}_{\ell,i}$, is the number of spikes in
channel $j$ for the $i^{\rm th}$ burst that had a leader $\ell$. We take the
average of these vectors, normalized to norm 1:
$$
{\bf c}_\ell = \sum_{i} {\bf c}_{\ell,i} / N_\ell~,
$$
where $N_\ell$ is the number of bursts with the leader $\ell$. Investigation of
these vectors shows that they are practically indistinguishable to the eye.

However, a prediction can be made for every burst whose leader is one of the
$\ell$ above, as follows: If $\bf C$ is the (normalized) vector corresponding
to the spike count in one specific measured burst, we form the scalar products
${\bf C}\cdot {\bf c}_\ell$ and predict that the leader is that $\ell$ for
which this quantity is largest. We do this for every burst. But we also know
who was actually the leader of the burst $\bf C$, say $\ell'$ and we therefore
get a matrix $A_{\ell',\ell}$ of counts of predictions $\ell$ for the case when
$\ell'$ was the actual leader. In a perfect network, the matrix $A$ would be
diagonal, since the prediction would never be wrong. This is too much to expect
for the noisy and untrained neural network we are studying, but the data show
clearly a high correlation between the predicted and the true leader. In a
totally random situation, the result would be evenly distributed between the
$n$ leaders and the relative abundance would be $A_{\ell',\ell}/\sum_\ell
A_{\ell',\ell}=1/n$. Thus we show in Table \ref{tab:fourbyfour} the quantities
$n A_{\ell',\ell}/\sum_\ell A_{\ell',\ell}$, and a coefficient $>1$ indicates a
positive correlation of the prediction.


\begin{table}[!h] 
\begin{center}
\begin{tabular}{|c|r|r|r|r|r|}
    &    22  &     32  &     44  &     55  &     57    \\
\hline
{\bf 22}  &   {\color{red}2.43\color{black}} &  0.19 &  1.08 &  0.87  & 0.43 \\
{\bf 32} &   1.24 &  0.99  & 1.59  & 0.73  & 0.45 \\
{\bf 44} &   1.41 &  0.47  & {\color{red}2.29\color{black}}  & 0.50 &  0.33 \\
{\bf 55} &   1.18 &  0.60 &  0.71  & {\color{red}1.78\color{black}} &  0.74 \\
{\bf 57} &   0.78 &  0.30 &  0.54  & 0.89 &  {\color{red}2.49\color{black}} \\
\end{tabular}
\quad
\begin{tabular}{|c|r|r|r|r|}
     &  12     & 19    &     40&       60 \\
\hline
{\bf 12} &   {\color{red}3.28\color{black}} & 0.29 & 0.20 &  0.24  \\
{\bf 19} &   0.67 & {\color{red}2.40\color{black}} & 0.61 &  0.32  \\
{\bf 40} &   0.86 & 0.16 & {\color{red}2.40\color{black}} &  0.59  \\
{\bf 60} &   1.80 & 0.26 & 1.21 &  0.73  \\
\end{tabular}
\end{center}
\caption{\small Two examples of the correlations between the leader (left
column) and the type of burst (top row). The left matrix is for superepoch 2 of
the \textit{CF} culture and the right is for superepoch 2 of a \textit{DF}
culture. Note the much higher scores on the diagonal (values higher than the
average by more than one standard deviation are colored red), as well as a
small uncertainty for the pair of close-by electrodes (22 and 32) in the left
matrix.} \label{tab:fourbyfour}
\end{table}

\begin{figure}[!h]
     \begin{center}
\includegraphics[width=0.66\textwidth,angle=0]{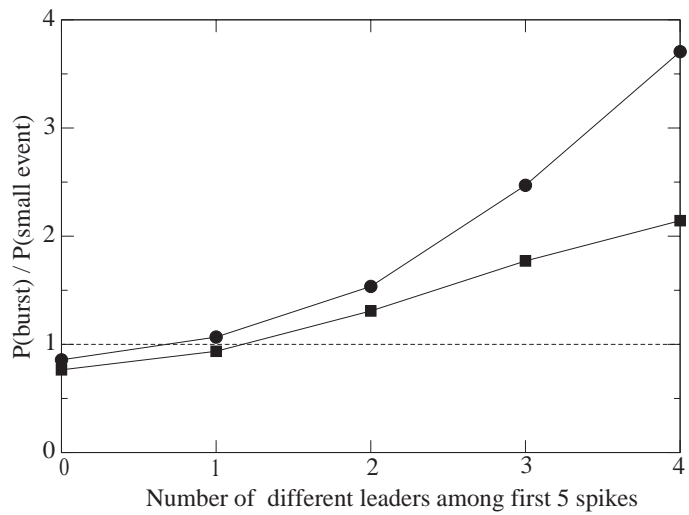} %
\caption{Prediction of burst initiation probability from the occurrence of
leaders in the first spikes in the pre-burst or small event. The ratio of
probabilities between burst and small event increases as a function of the
number of leaders in the 5 first spikes in the pre-burst. Results are given for
superepoch 1 (squares) and superepoch 2 (circles) of \textit{CF}. }
       \label{f:sme_prediction}
     \end{center}
   \end{figure}

A second test for the relation of leaders and bursts is done by another
prediction we make regarding the activity that follows a slow initiation. This
activity may increase to become a burst, or die away as a small event. Our
prediction is based on the number of leader electrodes that participate in the
first spikes. It turns out that when more leaders are active in the first few
spikes, the ratio between the probability of the activity to become a burst to
the probability to become a small event increases significantly
(\fref{f:sme_prediction}).

\section{Discussion}
The presence of leaders is an interesting fact, and we now analyze some of the
consequences and problems related to it. We have seen that the leaders are
first of all localized and stable, to which we now add the assumption that they
are homogeneously dispersed throughout the whole culture, although we have not
measured this. This assumption seems very reasonable, since leaders are
observed in all three preparations, and the location of the electrodes is not
expected to be special in any way. Although the electrodes cover an area that
is only a small part of the culture (typically reporting on less than $0.1\%$
of the neurons), they almost invariably succeed in picking up some leaders.

Two observations make us conjecture that `leaders' not only report on the
burst, but are active in causing it. The first is the locality of activity in
the pre-burst, indicating a recruitment process, \cite{YvonSpont}, in other
words transfer of information from the leader to its surrounding neurons prior
to the transition of the whole network to a full burst. This is reminiscent of
a similar recruitment process around BIZ's in 1D networks, in which transition
from low amplitude, localized activity to higher amplitude, fast propagating
activity was observed \cite{feinerman2005}, and that is related to the buildup
of synchrony in neural firing \cite{jacobi2007}. If the bursts were coming from
outside the monitored area then we would expect them to appear abruptly, with
no pre-burst and no gradual growth like the one we observe. So, although we
only observe a small part of the culture, leaders are clearly identified in the
field of observation for most of the bursts, indicating that the burst
generation must have a local component in its genesis.

The second observation is the predictive power of the leaders as to the
`character' of the burst. Several groups have reported on the existence of a
limited number of firing patterns in the bursts \cite{MaedaGeneration1995,
VanPeltDynamics, segev2004}. We have seen additionally that the spiking pattern
of each electrode within the burst provides a signature that suffices to
statistically identify its leader. Similarly, the firing pattern of leaders in
the first few spikes of the small event or the pre-burst allows a prediction as
to whether a burst will form or not.

We conclude from this that a relationship must exist between a burst
and local leaders, emphasizing again that this is by no means just a
statistical
abundance of spiking by the leaders, but a sign of signaling between
the leader, the exterior parts of the sample, and the burst generation.

What then do these leaders have as input that makes them fire first?
The dominant signal inputting on a given neuron is usually thought to
arise from the firing of other neurons rather than from environmental
cues such as fluctuations in the amount of ions in the surrounding
fluid. The reason that the leaders are first to fire may be that they
have the most inputs, or that they are more sensitive to
excitation. Recall however, that they do not fire particularly often.

There are three scenarios for how the leaders are excited. In the simplest case
one leader in the whole culture is excited by a localized low-level background
activity. By becoming very active it will recruit first its nearby environment
and then gradually the rest of the network. This scenario is rejected within
our picture: we have argued that several leaders are being excited
concurrently, otherwise we would not have seen leaders in the small region in
which we measure. The second scenario involves a number of leaders being
excited simultaneously by a widespread low-level noise that can activate many
of the leaders in parallel. In this case they each excite a region around them,
and these coalesce at random as the whole network bursts. The argument against
this scenario is that we would not expect then a variability in the signature
of the burst, since all bursts would evolve essentially in the same route, with
the leaders being excited first and then recruiting their respective
neighborhoods.

The third, most tentative yet most attractive scenario posits the existence of
an underlying `sub-network' connecting potential leaders. Their activation is
triggered by the noise, but here a small amount of background firing can
suffice to set off the burst since the leaders first communicate amongst
themselves and only then initiate the activity of other leaders. Variations in
the noise that initially sets off the network could then cause variation in the
response of the leader sub-network, effecting the parts of it that are
recruited. This in turn reflects on the details and signature of the burst.
This scenario is particularly intriguing since it implies a stochastic
percolation behavior in the sub-network of the leaders which, once they have
fired, act as nucleation sites for the whole network. Perhaps one could answer
the questions raised in this discussion experimentally, first measuring
passively and identifying leaders and then intervening by selectively
suppressing the activity of the neurons near those electrodes with leader
properties.

\section*{Acknowledgments}
We thank Steve Potter for making the data from his lab available to us. EM and
SJ thank Menahem Segal for advice, and SM thanks Vladimir and Elleonora Lyakhov
for technical assistance. JPE and CZ were supported by the Fonds National
Suisse, SM by a grant from the Israel Science Foundation, EM and SJ by grants
from the Israel Science Foundation, the Minerva Foundation (Germany) and the
Clore Center for Biological Physics. JPE and CZ are grateful for the
hospitality of the Einstein Center.

\section*{Bibliography}
\bibliographystyle{unsrt}
\bibliography{biblio}\addcontentsline{toc}{section}{Bibliography}

\pagebreak
\section*{Figures and captions}
\begin{figure}[!h]
\begin{center}
\includegraphics[width=0.54\textwidth,angle=0]{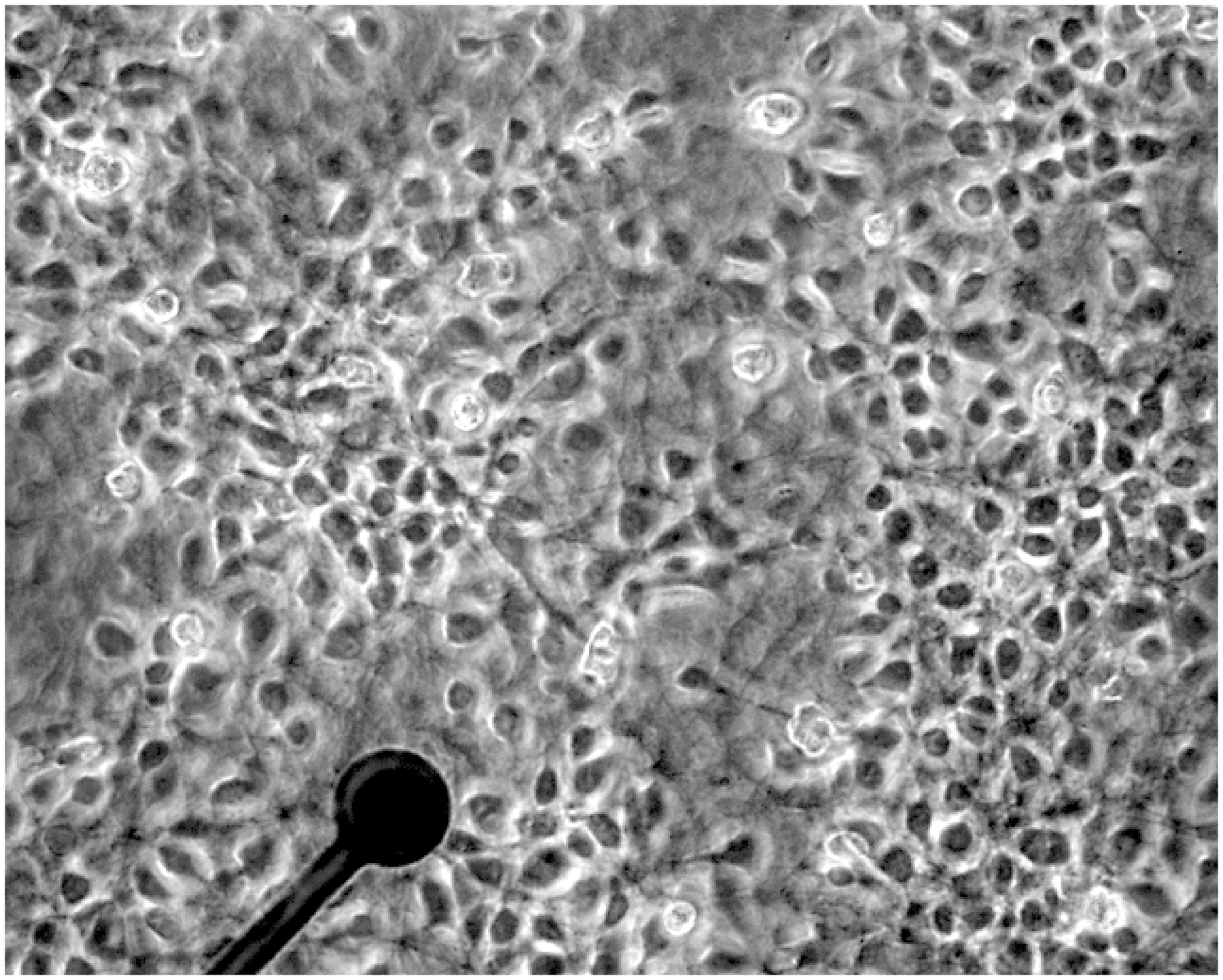}
\includegraphics[width=0.6\textwidth,angle=0]{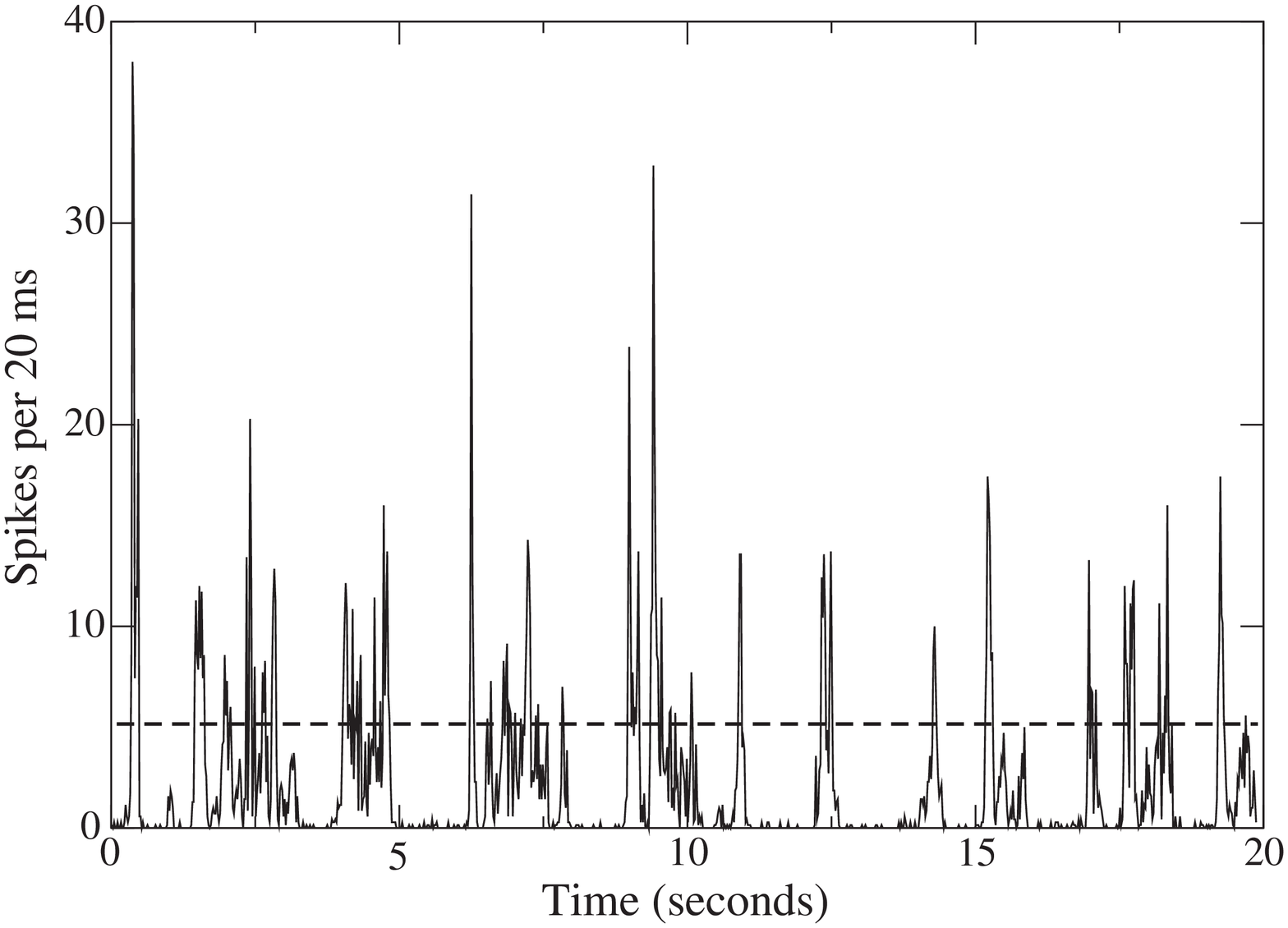} 
\caption{a. Two-dimensional rat hippocampal neuron culture
(\textit{DF}) growing near one of the 60 measuring electrodes of the
MEA. Electrode diameter is $30\mu m$. Phase micrograph taken using
Zeiss Axiovert 135TV. b. Example of spontaneous bursting activity
from a \textit{DF} culture. The number of spikes in 20ms bins is
drawn as a function of time. Burst detection threshold is 10 spikes
in 40ms (dashed line). } \label{f:two-dim-interval-drawing}
\end{center}
\end{figure}

\end{document}